\documentstyle[11pt,newpasp,rotcapt,twoside,psfig]{article}
\index{X-rays}
\index{radio}

\index{X-ray binaries!Cir~X-1}
\index{X-ray binaries!SS~433}
\index{X-ray binaries!associations with supernova remnants}

\index{anomalous X-ray pulsars!associations with supernova remnants}
\index{soft gamma-ray repeaters!associations with supernova remnants}

\index{radio-quiet neutron stars!associations with supernova remnants}
\index{radio-quiet neutron stars!1E~1207.4--5209}

\index{pulsar wind nebulae!G11.2--0.3}
\index{pulsar wind nebulae!G5.27--0.90}
\index{pulsar wind nebulae!Duck}
\index{pulsar wind nebulae!3C~58}
\index{pulsar wind nebulae!G130.7+3.1}
\index{pulsar wind nebulae!SN~1181}
\index{pulsar wind nebulae!Kes~75}
\index{pulsar wind nebulae!G29.7--0.3}
\index{pulsar wind nebulae!G292.0+1.8}
\index{pulsar wind nebulae!MSH~11--5{\em 4}}

\index{neutron stars!statistics}

\index{pulsars!Geminga}
\index{pulsars!J0631+1746}
\index{pulsars!J1124--5916}
\index{pulsars!B1757--24}
\index{pulsars!J0205+6449}
\index{pulsars!J1119--6127}
\index{pulsars!J1811--1925}
\index{pulsars!J1846--0258}
\index{pulsars!Duck}
\index{pulsars!statistics}
\index{pulsars!initial periods}
\index{pulsars!velocities}
\index{pulsars!ages}
\index{pulsars!associations with supernova remnants}
\index{pulsars!displacement from center of supernova remnants}

\index{supernova remnants!W50}
\index{supernova remnants!G39.7--2.0}
\index{supernova remnants!G321.9--0.3}
\index{supernova remnants!G292.0+1.8}
\index{supernova remnants!MSH~11--5{\em 4}}
\index{supernova remnants!G296.5+10.0}
\index{supernova remnants!PKS~1209--51/52}
\index{supernova remnants!Kes~75}
\index{supernova remnants!G29.7--0.3}
\index{supernova remnants!3C~58}
\index{supernova remnants!G130.7+3.1}
\index{supernova remnants!G292.2--0.5}
\index{supernova remnants!SN~1181}
\index{supernova remnants!G11.2--0.3}
\index{supernova remnants!G5.4--1.2}
\index{supernova remnants!Duck}
\index{supernova remnants!associations with anomalous X-ray pulsars}
\index{supernova remnants!associations with pulsars}
\index{supernova remnants!associations with radio-quiet neutron stars}
\index{supernova remnants!associations with X-ray binaries}
\index{supernova remnants!associations with soft gamma-ray repeaters}
\index{supernova remnants!statistics}

\def\lapp{\ifmmode\stackrel{<}{_{\sim}}\else$\stackrel{<}{_{\sim}}$\fi}
\def\gapp{\ifmmode\stackrel{>}{_{\sim}}\else$\stackrel{>}{_{\sim}}$\fi}

\def\edcomment#1{\iffalse\marginpar{\raggedright\sl#1\/}\else\relax\fi}
\reversemarginpar

\begin{document}
\title{Constraining the Birth Events of Neutron Stars}
\author{V. M. Kaspi}
\affil{McGill University, Rutherford Physics Building, 3600 University Street, Montreal, QC H3A 2T8, Canada}
\affil{Department of Physics and Center for Space Research, Massachusetts Institute of Technology, 70 Vassar Street, Cambridge, MA 02139, USA}

\author{D. J. Helfand}
\affil{Department of Astronomy, Columbia University, 550 West 120th St, 
Mail Code 5233, New York, NY 10027, USA}

\begin{abstract}
The prescient remark by Baade and Zwicky that supernovae beget neutron stars
did little to prepare us for the remarkable variety of observational
manifestations such objects display. Indeed, during the first thirty years
of the empirical study of neutron stars, only a handful were found to be
associated with the remnants of exploded stars. But recent X-ray and radio
observations have gone a long way toward justifying the theoretical link between
supernovae and neutron stars, and have revealed the wide range
of properties with which newborn compact remnants are endowed. We review
here our current state of knowledge regarding neutron star-supernova remnant
associations, pointing out the pitfalls and the promise which such links hold.
We discuss work on the ranges of neutron star velocities, initial spin periods, 
and magnetic field strengths, as well as on the prevalence of pulsar wind 
nebulae. The slots in neutron star demography held by AXPs, SGRs, radio-quiet
neutron stars, and other denizens of the zoo are considered. We also present an
attempt at a comprehensive census of neutron star-remnant associations and
discuss the selection effects militating against finding more such 
relationships. We conclude that there is no pressing need to invoke large
black hole or silent neutron star populations, and that the years ahead hold
great promise for producing a more complete understanding of neutron star
birth parameters and their subsequent evolution.

\end{abstract}

\section{Introduction}

The prediction by Baade \& Zwicky (1934) that neutron stars would form in 
supernova explosions ranks among the most prophetic
theoretical speculations in the history of astrophysics. An
important aspect of the matter these two illustrious astronomers
did not address was how such beasts could be observed.  Certainly the
properties of the rotation-powered pulsars discovered by Jocelyn Bell
were unanticipated before 1967; the thermal X-ray emission detected from
the surfaces of a handful of nearby young pulsars over a quarter of a
century later (e.g.\ Finley, \"Ogelman \& Kiziloglu 1992) was the first
properly predicted neutron-star observational characteristic.  But the
unanticipated reared its head in the interim, and the detection of soft
gamma-ray bursts from the supernova remnant N49 in the Large Magellanic
Cloud in 1979 (Mazets et al.\ 1979; Cline et al.\ 1982), as well as the
discovery, from a source in the remnant CTB~109, of slow X-ray pulsations 
that could be explained neither in terms of rotation-power nor in terms
of any conventional accretion mechanism (Fahlman \& Gregory 1981),
suggested that the words ``unexpected zoo'' may be applicable to the
observed young neutron star population.   This diversity of appearance
continues to grow: X-ray point sources with properties quite different from 
those described above include the variable point source in RCW~103
(Tuohy \& Garmire 1980; Gotthelf, Petre \& Vasisht 1999;
Garmire et al.\ 2000), the slow and X-ray variable pulsar AX
J1845$-$0258 in G29.6+0.1 (Vasisht et al.\ 2000), and the point source
in Cas~A, made famous by the spectacular first light {\em Chandra X-ray
Observatory}\ image (Tananbaum 1999).

It is perhaps ironic that in spite of the many possible observational
manifestations of young neutron stars, one of the greatest problems in
the field has been that most Galactic supernova remnants appear empty.
Observational selection effects must certainly play a role, but
ever-lingering is the fact that an unknown but possibly substantial
fraction of massive stars undergoing core collapse might produce black
holes rather than neutron stars.  The size of this fraction is
observationally poorly constrained.  Certainly the difficulty that
supernova modelers encounter in producing shock waves that can expel
the outer layers of a star following the bounce in core collapse does
little to dispel the concern that the black hole fraction
may be large (Burrows, Hayes \& Fryxell 1995; Liebend\"order et al.\
2001; Janka 2001).  Our ignorance of the neutron star equation of state
and hence the maximum stable neutron star mass adds to this confusion
(Baym \& Pethick 1979).  A radio pulsar is thought to be born between
every 60 and 330~yr in the Galaxy (Lyne et al.\ 1998).  This is in rough
agreement with the best-estimate core-collapse Galactic supernova rate
of one every $47 \pm 12$~yr (Tammann, L\"offler \& Schr\"oder 1994), although
a significant paucity of pulsars is suggested, and the uncertainties on
both quantities are large enough to leave plenty of room for
suspicion.

In this review, we discuss first the associations between SNRs and
classical rotation-powered neutron stars. (Henceforth, we refer to
the latter simply as ``pulsars''; non-rotation-powered pulsars, in
particular anomalous X-ray pulsars, will be specifically identified.)
Our emphasis is less on a critique of all proposed associations
(since significant reviews of this nature already exist --- see e.g.\
Kaspi 1998, 2000; Manchester 1998) and more on what fundamental
astrophysical insights such associations provide.  Indeed,
pulsar/supernova remnant associations have been heralded as, in
principle, constraining the neutron star initial spin period, velocity
and magnetic field distributions, all important for the physics of core
collapse.  In addition, dating pulsars is valuable in constraining
the equation of state by testing cooling models (e.g.\ Umeda et al.\
1994) --- associations with supernova remnants can in principle offer
independent age determinations.

We subsequently consider associations between supernova remnants and
non-rotation-powered neutron stars as well as compact remnants of an
uncertain nature, not merely for completeness, but as a precursor to
the final section, in which we present an attempt at a synthesis of all
associations.  The goal here is to address the larger issue of
core-collapse-remnant demographics.  Similar attempts have been made in
the past (e.g.\ Helfand \& Becker 1984; Helfand 1998), although the new
observational capabilities in the high energy regime represented by
{\em Chandra}\ and {\em XMM-Newton}\ have led, in the last two years,
to a dramatic acceleration in our identification of the compact
remnants of supernova explosions. Indeed, we conclude here that current
data are consistent with the notion that the majority of core-collapse
events in the Galaxy {\it do} leave neutron star remnants.

\section{Rotation-Powered Pulsars and Supernova Remnants}

\subsection{Making Associations}

Associations between pulsars and remnants are generally made from the
positional coincidence of an apparently young pulsar with a remnant
that has no other more plausible association.  Ideally, distance estimates
to the two sources should agree, although neither is generally determined
with high precision; similarly for rotation measures when they are
available.  Given the independently observed high space velocities of pulsars
(Lyne \& Lorimer 1994),
significant offsets of the neutron star from its birth place should be expected,
particularly for older associations (Gaensler \& Johnston 1995).  These
provide a possible test of an association if a proper motion measurement
can be made (although arguments for alternative causes for such offsets are
presented by Gvaramadze, these proceedings).

Here we highlight several recent results concerning rotation-powered
pulsar/supernova remnant associations of particular relevance to the field as a
whole.

\subsection{Inferring Neutron Star Space Velocities: A Dangerous Hobby}

PSR B1757$-$24 has long been held up as the poster pulsar for high space
velocities inferred from supernova remnant associations, in this case the shell
remnant G5.4$-$1.2 (Frail \& Kulkarni 1991; Manchester et
al.\ 1991).  The pulsar position, just outside the remnant boundary, suggests that
the neutron star has overtaken the shell, which, unlike the pulsar, suffers
deceleration from swept up interstellar material.  The morphology of
a small, flat-spectrum nebula protruding from G5.4$-$1.2, G5.27$-$0.90,
strongly suggests ram-pressure confinement of the pulsar wind, lending
significant support to this picture.  Additional evidence for the
pulsar having passed through the shell is the fact that the pulsar side
of the remnant is brightest, and that the shell spectral index decreases
monotonically along the shell with distance from the pulsar (Frail, Kassim \&
Weiler 1994).

The pulsar characteristic age, $\tau_c \equiv P/2\dot{P}$ (where $P$ is
the spin period and $\dot{P}$ its rate of change) is 16\,000~yr for PSR
B1757$-$24.  Of course, $\tau_c$ is only a good age estimator if the
pulsar's initial spin period is much shorter than that observed
today, and if simple magnetic dipole braking  is applicable throughout
the life of the pulsar (see Eq. 1 below).  
Nevertheless, using $\tau_c$ as a reasonable
estimate of the true age, and assuming the pulsar was born at the shell's
geometric center, implies a pulsar proper motion of
$\sim70$~mas~yr$^{-1}$, or a transverse velocity of
$\sim1800$~km~s$^{-1}$ for a distance of 5~kpc.  This would make
PSR~B1757--24 the fastest pulsar known.

However, Gaensler \& Frail (2000) measured the proper
motion to be $< 25$~mas~yr$^{-1}$ at the 5$\sigma$ level.  They
interpret this result as suggesting that the system is much older than
$\tau_c$ implies; they argue that the pulsar's braking index $n$,
assumed to be 3 in standard vacuum magnetic dipole braking, must be
less than 1.83, and possibly even less than 1.33.  Evidence exists for
such low braking indices (e.g.\ Lyne et al.\ 1996; Zhang et al.\ 2001).
Unfortunately X-ray observations of the remnant, which in principle
could help independently constrain its age via the shock temperature,
yield only an unconstraining upper limit, given the large absorption
toward the source (see Kaspi et al., these proceedings).

Of course, the chance superposition of the pulsar and
G5.4$-$1.2 cannot be ruled out, nor can an offset between the pulsar
birth place and the shell center because of the space velocity of the
progenitor (see Gvaramadze, these proceedings).  In any event, the
pulsar/supernova remnant picture in general is clearly complicated by
the surprisingly low proper motion for PSR~B1757--24.  Inferring
velocity estimates from possible associations appears to be a dangerous
hobby.

Unaddressed by Gaensler \& Frail (2000) 
is the interesting question of why a pulsar of
apparently average velocity should exhibit so spectacular and unique a
radio wind nebula.  Recent {\em Chandra}\ observations of the system
have detected X-ray emission coincident with the radio nebula,
supporting the ram-pressure confined wind interpretation (see Kaspi et
al., these proceedings).  One possibility is that a condition for the formation
of a bright ram-pressure confined pulsar wind nebula is strong rear
confinement, as offered by the high pressure interior of a supernova
remnant shell.

\subsection{Neutron Star Initial Spin Periods}

The age of a pulsar, $\tau$, can be calculated, assuming a constant braking
index, $n$, from birth, by integrating the standard spin-down expression
$\dot{\nu} \propto \nu^n$ (where $\nu \equiv 1/P$) to yield
\begin{equation}
\tau = \frac{P}{(n-1)\dot{P}} \left[ 1 - \left(\frac{P_0}{P}\right)^n \right],
\end{equation}
where $P_0$ is the spin period at birth.  Clearly, for $n=3$ and $P_0 \ll P$,
one recovers the conventional definition of characteristic age.
Only the Crab pulsar has, until now, had a solid initial spin
determination, $P_0 = 19$~ms,
thanks to the fact that $n$ has been measured and the 
year of the system's birth is known
independently.  Interesting constraints on other initial spin periods
exist. For PSR~J0537$-$6910, $P=16$~ms.  Given that its host remnant
N157B has an approximate age of 5~kyr, this suggests $P_0 \ll 16$~ms for
$n=3$, and $P_0 < 10$~ms for $n \lapp 2$ (Marshall et al.\ 1998).  PSR
B1951+32 in CTB~80 has $\tau_c = 107$~kyr, comparable to the dynamical
shell age (Koo et al.\ 1990); this suggests  $P_0 \ll 39$~ms for $n=3$.
Other systems provide higher and therefore less interesting upper
limits on initial spin period.  A variety of pulsar population
synthesis models have therefore assumed a mean value of $P_0 \simeq
20$~ms in their simulations (e.g.\ Narayan \& Ostriker 1990; Lorimer et
al.\ 1993; Cheng \& Zhang 1998; McLaughlin \& Cordes 2000).

G11.2$-$0.3 is a bright, highly symmetrical circular shell supernova
remnant detected both at radio and X-ray wavelengths (Downes 1984; Morsi
\& Reich 1987; Green et al.\ 1988).  Its age, inferred from its
morphology and X-ray spectrum (Aoki 1995), has been estimated to be
$\sim2000$~yr.  A possible association with a ``guest star'' reported
by Chinese observers in 386 A.D. also suggests youth (Clark \&
Stephenson 1977).

However, Torii et al.\ (1999) reported that the 65-ms X-ray pulsar 
PSR~J1811--1925 discovered within the remnant (Torii et al.\ 1997) has
$\tau_c = 24$~kyr.  Assuming an association, they reconcile the age
discrepancy by arguing that either $P_0 \simeq 65$~ms, or
that the pulsar's braking index is an order of magnitude larger than
the canonical value of 3.  {\em Chandra}\ X-ray observations of the
system pinpointed the pulsar position as lying very close to the remnant
center, supporting its inferred youth (Kaspi et al.\ 2001b).  Given that, of the
five solid braking index measurements (Lyne, Pritchard \& Smith
1988; Kaspi et al.\ 1994; Lyne et al.\ 1996; Camilo et al.\ 2000; Zhang et
al.\ 2001), not one has been reported as greater than 3, a value of $\sim30$
for PSR J1811$-$1925 seems hard to swallow, making $P_0 \simeq 62$~ms
virtually inescapable.  Nevertheless, phase coherent timing
observations to verify this are planned.  Furthermore, it should be
possible to measure directly the expansion of the remnant in order to
confirm its youth.  Thus, it seems likely that the G11.2$-$0.3 system
will provide only the second solid determination of a birth spin
period for a neutron star; this system offers further proof that
characteristic ages should be treated with extreme caution,
particularly among apparently young pulsars.

Indeed, the latter point has broader implications.  Of all pulsars known
having $\tau_c < 100$~kyr, $\sim25$\% have $P<90$~ms.
The long birth spin period of PSR~J1811$-$1925 thus suggests that this
fraction of young pulsars could have characteristic ages that are gross
overestimates of their true age.  Of course, small braking indexes
could offset the discrepancy.

Recently, Murray et al.\ (2002) have discovered PSR~J0205+6449 in the
Crab-like remnant 3C~58.  The latter has long been thought to be
associated with the historical event recorded in 1181~A.D. (Clark
\& Stephenson 1977).  The 66-ms pulsar, however, has a characteristic
age 5400~yr, far in excess of that implied by the historical identification.
This led Murray et al.\ (2002) to suggest that, like the G11.2$-$0.3 system,
the initial spin period of PSR~J0205+6449 would have to be long (roughly 60~ms
for braking indices in the range 1.5--3).
However Bietenholz, Kassim \& Weiler (2001) have suggested, on the basis
of remnant expansion measurements, that the true remnant age is 
$\sim5000$~yr, in agreement with the pulsar characteristic age,
calling into question the hypothesized long initial spin period.

Spruit \& Phinney (1998) have suggested that there could be a
correlation between initial spin period and space velocity if the
supernova asymmetry that imparts a ``kick'' to the neutron star does so
in a single blow with an impulse that is not aligned with the neutron
star center-of-mass.  The G11.2$-$0.3 pulsar has a lower transverse
velocity ($v_t < 110$~km~s$^{-1}$) than those of the Crab pulsar ($v_t
= 150$~km~s$^{-1}$; Caraveo \& Mignani 1999), PSR~B1951+32 ($v_t \simeq
300$~km~s$^{-1}$; Fruchter et al.\ 1988) and the N157B pulsar ($v_t \sim
600$~km~s$^{-1}$; Wang \& Gotthelf 1998). 
All three certainly have shorter initial spin periods, however, arguing 
against a single, off-center impulse scenario. A proper motion for
PSR~J0205+6449 has yet to be measured, but is clearly of interest.

\subsection{The Observability of Plerions and the Empty Shell Problem}

PSR~J1119$-$6127 is a young radio pulsar found in the Parkes multibeam
survey (Camilo et al.\ 2000; see contributions by Manchester et al.\ and
Crawford et al., these
proceedings).  The pulsar's period $P=408$~ms and its rate of slow-down
imply $\tau_c = 1700$~yr, a spin-down luminosity $\dot{E} = 2.3
\times 10^{36}$~erg~s$^{-1}$, and, most notably, a very large implied
surface dipolar magnetic field, $B= 4.1 \times 10^{13}$~G.  The young
age of the system is assured, as Camilo et al.\ (2000) have also measured the
braking index $n=2.91 \pm 0.05$.

Deep Australia Telescope Compact Array radio observations
revealed a shell of emission with a non-thermal spectrum, centered on
the pulsar position (Crawford et al.\ 2001).  They name this
supernova remnant shell G292.2$-$0.5, and note that its size supports
the youth of the system.  X-ray observations also detect the remnant,
and are suggestive of an unusually hard X-ray spectrum
(Pivovaroff et al.\ 2001).

Although there is tentative evidence for possible unpulsed X-ray emission 
associated with the pulsar, likely a synchrotron nebula, there is a noteworthy
absence of any radio nebula.  This makes the PSR
J1119$-$6127 / G292.2$-$0.5 system an essentially unique example of an
energetic, very young pulsar in a pure radio shell remnant.  (PSR~0538+2817 in
S147 [see Table 1] has estimated age 620~kyr and 
$\dot{E} = 5 \times 10^{34}$~erg~s$^{-1}$, so is not expected to have
an observable synchrotron nebula.)  Crawford et al.\ (2001)
show, using a detailed model of synchrotron nebular evolution (Reynolds
\& Chevalier 1984), that the absence of an observed radio nebula could
result from strong losses suffered by the source early in its evolution,
as a consequence of its high magnetic field.  This system thus suggests that many
empty shell supernova remnants may harbor rotation-powered pulsars that
produce no observable plerions, a notion that could ameliorate the
``empty shell'' problem.

However, the young PSR~J1846$-$0258, recently discovered in the
supernova remnant Kes~75 also has a very high magnetic field, $4.8
\times 10^{13}$~G (Gotthelf et al.\ 2000), yet appears to power a radio
synchrotron nebula whose flux is significantly underpredicted by the same
line of reasoning used for PSR~J1119$-$6127 (Crawford et al.\ 2001).
A braking index for PSR~J1846$-$0258 has yet to be measured however,
so its true age is not known.  If the Reynolds \& Chevalier (1984)
model can describe this system as well, then the pulsar should be
significantly younger than PSR~J1119$-$6127, in agreement with 
the claim of Gotthelf et al.\ (2000)
that it is the youngest Galactic pulsar.  Accordingly, $n \approx 3$
is expected.  On the other hand, as discussed in \S~2.2, low braking 
indexes could be common.  For example, for $n=1.8$, PSR~J1846$-$0258
would have age $\sim$1800~yr, larger than that of PSR~J1119$-$6127, and
hence in disagreement with the predictions of Reynolds \& Chevalier
(1984).

\section{Alternative Neutron Star Manifestations}

\subsection{Anomalous X-ray Pulsars and Soft Gamma Repeaters}

Anomalous X-ray Pulsars (AXPs) are a small class of pulsating X-ray
sources which cannot be powered by rotation and which are highly
unlikely to be powered by accretion.  There are five confirmed members
of this class, as well as one AXP candidate.  For substantial
discussions on AXPs, see the contribution by Gavriil \&
Kaspi in these proceedings, as well as a review by Israel,
Mereghetti \& Stella (2002).
The Soft Gamma Repeaters (SGRs), of which there are
four confirmed examples and a possible fifth, are sources which
occasionally emit bursts of soft gamma-rays with an enormous fluence.
They also show AXP-like pulsations in quiescence.
Discussions of these sources have been presented by Hurley (2000).

Here we consider the AXPs and SGRs together because they bear a number
of striking similarities that strongly suggest they are
closely related (Thompson \& Duncan 1996).  In particular, both exhibit 
pulsations in the narrow
range of periods from 5~s to 12~s, and both show long-term, rapid
spin-down, with spin-down rates that imply short characteristic ages
and ultra-high magnetic fields (assuming simple dipole braking, which
may not be the case).  In addition, they share similar X-ray spectra in
quiescence, and have similar timing properties (Woods et al.\ 2000;
Kaplan et al.\ 2001; Kaspi et al.\ 2001a).

The magnetar model, proposed for both types of objects, posits that the
X-ray emission is powered by magnetic field decay, and that the soft 
gamma-ray bursts come from surface cracking and subsequent particle emission
arising from magnetic stresses in the neutron star crust (Thompson \& Duncan
1996).   A competing model for the nature of AXPs invokes
accretion from a supernova fall-back disk (Chatterjee, Hernquist \&
Narayan 1999).  However, the faintness of optical/IR counterparts
(Hulleman et al.\ 2000, 2001) and the recently announced detection of
optical pulsations from one AXP (Kern \& Martin 2001) are claimed
to rule out this scenario conclusively, although details are yet to emerge.

Among the primary motivations for the magnetar model was the
coincidence of the first discovered SGR, SGR 0526$-$66, with the LMC
supernova remnant N49 (Cline et al.\ 1982). Subsequent observations found the
burster to be significantly offset from the remnant center, though still well
within its projected boundary (Rothschild, Kulkarni \& Lingenfelter 1994;
Marsden et al.\ 1996).  Attempts have been made to associate the three
other well-localized SGRs with supernova remnants (Kulkarni et al.\ 1994;
Hurley et al.\ 1999; Woods et al.\ 1999) but none has withstood the test
of time convincingly (see Gaensler et al.\ 2001 for a review).  This
does not rule out SGRs being young neutron stars --- of the ten known rotation-powered
pulsars with characteristic ages under 10~kyr, two are coincident
with no obvious supernova remnant.  Finding associated remnants for SGRs
is further complicated by their location in complicated
regions of the Galactic plane associated with massive star
formation (Vrba et al.\ 2000).

Two of the five confirmed AXPs, plus the AXP candidate, are located
very near the centers of well established supernova remnants:
1E~2259+586 in CTB~109 (Fahlman \& Gregory 1981), 1E~1841$-$045 in
Kes~73 (Vasisht \& Gotthelf 1997), and the possible AXP AX~J1845$-$0258
in G29.6+0.1 (Gaensler, Gotthelf \& Vasisht 1999).  Although the age of 
1E~2259+586 is much larger
than that of CTB~109 (225~kyr versus $\sim20$~kyr), this could be due to
torques from Alfv\`en wave and particle emission (Thompson \& Blaes 1998),
or because the magnetic field decays on a timescale of $\sim 10^4$ yr (Colpi, 
Geppert \& Page 2000). In addition, possible spin-up events have been reported (e.g.\ Baykal \& Swank 1996) suggesting that simple age estimates are
unreliable. However, recent phase-coherent observations have detected only
steady spin-down for this source over 4.5 yr (Gavriil \& Kaspi 2002).

Marsden et al.\ (2001) have claimed, on the basis of putative
associations with SNRs, that the regions in which AXPs and SGRs are
embedded have higher than average ambient densities, allowing for
interesting mass-infall rates, and hence accretion power. In our view,
this claim has been effectively rebutted by Gaensler et al.\ (2001); in
light of the other arguments against accretion, we regard this power
source as unlikely.  On the other hand, Gaensler et al.\ (2001) argue, on the
basis of associations (and the lack thereof) with supernova remnants,
that SGRs are a high-velocity population, and that, in spite of the
many similarities described above, likely represent a different class
of sources from the AXPs.  However interesting such speculation may be,
we consider the conclusion premature given that (i) some AXPs (and
young pulsars) are unassociated with remnants, (ii) it may be harder to
find associated remnants for SGRs because of their complex
environments, (iii) the numbers of both source types are small, and
(iv) there are evident risks inherent in {\it a posteriori} probability
estimates.

\subsection{``Radio-Quiet'' Neutron Stars}

The epithet ``radio-quiet'' has been applied to a number of neutron
stars and neutron-star candidates in the last few years (see Brazier \&
Johnston 1999 for a recent review). We begin by
making clear the term's anthropocentric definition: objects from which
we astronomers, on Earth, have yet to detect radiation in the
wavelength range 1 m to 10 cm.  A variety of reasons can be put forth
to explain this failure: (1) the pulsar's radio radiation is beamed (an
uncontroversial assumption) and our line of sight does not intercept
the beam; (2) the pulsar fails to emit any low frequency radiation
because the radio emission mechanism is (a) shorted out by accretion, (b)
inoperative because the magnetic field is too high (Baring \& Harding
2001), (c) inoperative because the polar cap potential drop is too low (Chen 
\& Ruderman 1993; Hibschman \& Arons 2001), or
(d) inoperative for reasons we don't understand; (3) the radio emission
is below current search thresholds; or, (4) the X-ray point source in
question is not a neutron star.  Each of the sources in this category discussed
below can be explained by at least one of the above possibilities.

Geminga (PSR~J0631+1746) is the first, and perhaps most famous case of
a radio-quiet pulsar. The second brightest object in the 100-MeV sky,
the source was finally identified as a rotation-powered pulsar when
X-ray pulses were detected by {\em ROSAT}\ (Halpern \& Holt 1992). The
source's $P$, $\dot{P}$, ratio of gamma-ray luminosity to spin down
energy loss rate, space velocity, and other properties are typical of
intermediate-age ($\sim10^{5.5}$ yr) radio pulsars. Its proximity and
radio flux density upper limit place its observable radio luminosity
(from the 320 MHz flux density) a factor of ten below that of the least
luminous pulsar known, and a factor of $\sim 300$ below the weakest
pulsars of comparable age and $\dot E$ (McLaughlin et al.\ 1999). While other physical
mechanisms may be responsible for the absence of radio emission (e.g.\
quenching of the low-altitude radio particle accelerator by pairs
streaming back toward the surface from the outer-gap gamma-ray emission
region; Halpern \& Ruderman 1993), simple beaming remains a viable
explanation.

The X-ray point source near the center of the southern supernova
remnant PKS~1209--51/52 (G296.5+10.0) has also been known for more than
twenty years (Helfand and Becker 1984), and reasonably stringent radio
upper limits have been obtained for any pulsed flux from the source
(Kaspi et al. 1996).  Recently, Zavlin et al. (2000) have detected
X-ray pulsations with a period of 0.424 s. While a period derivative
has yet to be determined, the X-ray properties of the source mark it as
-- dare we say it -- ``prototypical of an emerging class'' of X-ray
sources in supernova remnants. It has an X-ray luminosity of $\sim
10^{33.5}$ erg s$^{-1}$ and a soft X-ray spectrum which, if parameterized
as a black body, yields a temperature of 0.25 keV. The original notion
that sources such as this were cooling neutron stars has been challenged by
the fact that the inferred emitting radius is only 1.1 km. However, when
complex neutron star atmospheres are used to fit the spectrum (and
lower the inferred temperature), the emitting radius can be raised
to a plausible value (Zavlin, Pavlov, and Tr\"umper 1998). There is, as yet,
no complete model that both incorporates observed small pulsed fraction
($\sim 8\%$) and the spectral fits (Pavlov, this volume).
The radio upper limit translates to a luminosity significantly
below that of most other young sources, but a few objects with
comparable periods have lower radio luminosities. The lack of a radio
synchrotron nebula suggests a comparison with PSR~J1119$-$6127 (\S 2.4)
which has a similar period, age, and surrounding shell-type supernova
remnant. If a high value of $\dot P$ is measured for this source, the
analogy will be all the more compelling.

Three other point-like X-ray sources inside SNR shells have very
similar X-ray properties to the PKS~1209--51/52 central source and, in common
with this source and Geminga, all lack evidence for surrounding pulsar wind
nebulae. These objects lie
within the confines of RCW~103 (Tuohy \& Garmire 1980), Puppis~A (Petre, Becker,
\& Winkler 1996), and Cas~A (Tananbaum 1999), the youngest known
remnant in the Galaxy. All have X-ray temperatures of 0.3--0.6~keV and
estimated X-ray luminosities within a factor of three of $10^{33.5}$~erg~s$^{-1}$, although the first of these has been reported to have
varied by more than a factor of ten on a timescale of months (Gotthelf,
Petre \& Vasisht 1999; Garmire et al.\ 2000). In addition, the
inferred blackbody radii for each source are 0.5 to 2 km (see
Chakrabarty at al.\  2001 for a detailed comparison of the X-ray
properties of these objects with those of other young neutron star
classes).  While it is possible to envision scenarios in which a small
hotspot rather than the whole neutron star is producing most of the
X-ray emission (e.g.\ as a consequence of chemical abundance gradients
in the stellar envelope [Pavlov et al.\ 2000] or as a result of
accretion onto the poles of a weakly magnetized star [Chakrabarty et
al.\ 2001]), such localized emission must, perforce, suffer rotational
modulation, a phenomenon yet to be observed in these three objects,
although deeper searches for pulsations continue.

A note of caution on the radio upper limits for these objects is
occasioned by the recent result from Camilo et al.\ (2002) announcing
the discovery of a radio pulsar in G292.0+1.8. They show that this
young ($<2000$ yr) pulsar has a radio luminosity of only $\sim 2$
mJy kpc$^2$, more than an order of magnitude below the least luminous
young pulsar previously known; this value is a factor of ten
below the upper limit on the Cas A X-ray point source (McLaughlin et al.\
2001), and is similar to the current upper limits for the recently discovered
rotation-powered pulsar in 3C~58 and the other three sources discussed here.

\subsection{Binaries}

The majority of all massive stars are in binaries. The preceding
discussion of a score of remnants
of such stars, however, has included only single systems.
The canonical explanation for the fact that only two
of over 1000 (unrecycled) pulsars has a binary companion
is that, either through the loss of more than half the mass
of the system, or as a consequence of an asymmetric kick
at the time of the explosion, most binaries are unbound by
a neutron star's progenitor supernova. Population
synthesis models of core-collapse explosions in binary
systems can, with some tweaking, do a reasonable job
of reproducing the observed population of high-mass X-ray binaries
containing neutron stars as well as the single pulsar population.

The detailed population synthesis by Portegies Zwart \& Verbunt
(1996) provides a useful example of such a calculation. They adopt
a core-collapse supernova rate of 1.6 per century and a population
in which 60\% of the massive stars are single and 40\% are in binaries.
They include the effects of primordial binary eccentricities and
mass transfer. In their models with no kick provided to the neutron star
at birth, they predict 1 in 8 young neutron stars should have a companion.
However, this scenario overproduces the number of Be star binaries and other
binary pulsar systems by a large factor. With a natal kick of 450 km s$^{-1}$,
the predicted number of neutron stars with a binary companion at birth falls
to 1 in 90 (and nearly 25\% of these are Her X-1 type systems which, the
authors acknowledge, are over-produced in all their models). This
binary birthrate of roughly one system per 5000 yr means we should
not expect more than half a dozen such systems in the Galaxy to
be associated with their progenitor supernova remnants.
In fact, we know of only two candidates for young neutron stars associated
with supernova remnants which do have binary companions. Despite
decades of study, however, both cases remain ambiguous.

Cir X-1 certainly contains a neutron star because it exhibits X-ray
bursts (Tennant, Fabian \& Shafer 1986), and is indubitably a binary
with a 16.6 day period  (Kaluzienski et al.\ 1976). However, it lies
$25^{\prime}$ from the center of G321.9$-$0.3, and requires a systemic
space velocity of $>580$ km s$^{-1}$ to explain its location. Tauris et
al.\ (1999) explore the implications of this association. They find a
minimum age for the remnant of 60\,000 yr and a required impulse to the
system of 500--1000 km s$^{-1}$.  While not energetically excluded,
this is uncomfortably large compared to the observed velocities of most
pulsars (i.e.\ had the neutron star alone received such a kick, its
velocity would have been well in excess of 1000 km s$^{-1}$). Thus,
despite some morphological evidence for the association (Stewart et al.\
1993; Fender et al.\ 1998), we regard the connection of the binary
system with the supernova remnant as less than certain.

SS~433 is the other candidate system. In this case the binary nature
of the source is also unambiguous, but opinion is divided as
to whether W50 is a supernova remnant (Kirshner \&
Chevalier 1980) or a wind-blown
bubble produced by the source's relativistic jets (K\"onigl 1983), and
whether the compact object is a neutron star or a black hole.
On the latter subject, opinion is divided, sometimes within the minds of
individual authors (cf.\ Zwitter \& Calvani 1989 and D'Odorico
et al.\ 1991). Long considered unique in the Galaxy for its
precessing relativistic jets, the recent discovery of the ``micro-quasars''
which show similar radio jets (Mirabel \& Rodr\'{\i}guez 1999) is relevant;
Safi-Harb, Durouchoux \& Petre (2001) have placed SS~433 in this class
which is apparently dominated by black hole accretors.

Even if one accepts both sources as examples of young binary neutron stars
associated with supernova remnants, they represent less than 5\%
of the confirmed young neutron stars discussed above and in \S4.
As noted earlier, this is to be expected if neutron stars receive a substantial
impulse at birth. As the observed population of pulsars in supernova remnants
becomes more complete and more fully characterized, it will be able to provide a
constraint on the range of kicks required to unbind most systems at the time of
the first supernova event. A corollary of this high binary destruction rate
is that many remnants should contain within their shells the discarded
binary companions. Finding such objects and determining their space velocities
would further constrain the kinematic consequences of pulsar birth.

\section{A Status Report on Associations}

\subsection{The Census}

Recognizing the variety of manifestations young neutron stars can exhibit and
the ambiguities (cited above) involved in making definitive associations, we
nonetheless attempt here a summary of our current state of knowledge. In Table
1, we provide a list of all remnants which at least two workers in the field
(the authors) can agree exhibit some sign of an associated young neutron
star. We utilize the standard SNR classifications of shell-like, composite
and Crab-like. We divide the table into four categories representing various
levels of certainty concerning the evidence: the detection of a young,
rotation-powered pulsar within the confines of the remnant, the detection of
an X-ray point source, the detection of a diffuse, non-thermal X-ray nebula indicating current particle acceleration accompanied by a radio synchrotron
nebula, and the detection of a radio synchrotron nebula alone lacking high
energy confirmation.

Seventeen years after the discovery of the first radio pulsar, only three
rotation-powered neutron stars where known to be associated with Galactic
supernova remnants (Helfand \& Becker 1984). Seventeen years after that
review, the number has grown to fifteen, with nearly half of these discovered
in the past few years. Two additional examples are found in the Large 
Magellanic Cloud. In all but two cases, the neutron star is surrounded by a radio and/or X-ray
pulsar wind nebula. Thus, in at least $\sim10\%$ of the cataloged
remnants in the Galaxy likely to have resulted from a core-collapse 
supernova\footnote{As argued below, $\sim50$ cataloged 
Galactic remnants had SN~Ia
progenitors.} (and a considerably larger fraction of the
younger remnants --- see below), we directly observe the ``canonical'' result:
a rapidly spinning, magnetized neutron star losing energy via a relativistic
wind which powers a surrounding nebula.  

Quantitatively, however, this population does not follow the canon
which posits the Crab pulsar as a prototype. We have known this
implicitly for three decades: there should be at least half a dozen
pulsars in the Galaxy within a factor of two of the Crab in age if we
adopt the Lyne et al.\ (1998) lower birthrate limit of 1 per 330 yr, and
more than a dozen for the best birthrate estimate. Yet the first X-ray
surveys of the sky showed no such objects of comparable luminosity.
Thus, the $\sim 19$~ms initial spin period, $\sim 10^{12.5}$~G field,
relatively low velocity ($\sim 150$ km s$^{-1}$), and confining
filaments are at least not {\em all}\ typical. Indeed, the observed
periods for the young neutron stars known today range over one and a
half orders of magnitude: from 16~ms for the source in N157B to 424~ms
for the X-ray pulsar in PKS~1209--51/52; even inferred spin periods at
birth have at least a spread of a factor of ten (\S 2.3).  Magnetic
field strengths are likewise broadly distributed, from $1 \times
10^{12}$~G for N157B to $50 \times 10^{12}$~G for the pulsar in Kes 75.
While the relationship among these observed parameters and the initial
periods and magnetic field strengths is not at all well understood, it
is likely that a considerable range in birth properties exists, and that
the Crab represents an unusual combination that renders it far more
conspicuous than most young neutron stars.

The second set of candidate young neutron stars contains the varied
collection of X-ray point sources associated with supernova remnants
(or putative remnants) discussed in \S3. Two are AXPs, one is a soft
gamma-ray repeater, and at least two are X-ray binaries (SS~433 --- not
obviously a neutron star --- and Cir~X-1). The remainder are X-ray
point sources of an ambiguous nature, although those with known,
pulsar-like periods (PKS~1209--51/52) or with surrounding X-ray and
radio synchrotron nebulae (e.g.\ G54.1+0.3, G0.9+0.1) are highly likely
to be confirmed as neutron stars in the near future.

Given the short ($\sim 1$ to 100 yr) lifetime of
X-ray-emitting synchrotron electrons in the expected nebular magnetic fields
($\sim 10^{-4}$~G), a centrally peaked diffuse X-ray source with a power-law
X-ray spectrum inside a supernova remnant provides a compelling argument for the
presence of a young pulsar even when no X-ray or radio point source is detected.
Nine additional remnants exhibit such evidence. Deeper imaging observations with 
{\em Chandra}, and timing observations with {\em XMM-Newton}\ and radio
telescopes may soon reveal pulsars within them, just as recent
observations of 3C~58 (Murray et al.\ 2002), G106.6+2.9 (Halpern et al.\
2001), and G292.0+1.8 (Camilo et al.\  2002) have promoted these objects
to the confirmed category. Less certain candidates are found among the
eight objects which appear to have a flat-spectrum synchrotron
component in their radio spectra. For many of these, the X-ray upper
limits are weak or non-existent; future observations could readily
confirm or refute the presence of an active young pulsar.  Nonetheless,
the final total for cataloged remnants with some reasonably sound
evidence for a contemporaneous young neutron star is 48 (plus three in
the LMC) --- a substantial improvement over the handful known only a
decade ago.

\subsection{Selection Effects}

The rapid rise in the last few years in the number of compact-object/SNR 
associations demonstrates that previous searches were simply lacking the
sensitivity necessary to detect the various manifestations of young neutron 
stars. Yet an examination of the properties of the remnants in Table 1
shows clearly that our current sample remains substantially incomplete.
In Table 2, we display the Galactic longitude- and latitude-dependence,
as well as the angular size distribution, for the remnants exhibiting evidence
for a compact object. The results demonstrate that striking selection
effects remain in the sample. For example, only 10\% (7/69) of the remnants 
within $20^{\circ}$ of the Galactic Center show such evidence, whereas the
fraction in the sector $180^{\circ}<l<300^{\circ}$ is 33\%. Higher X-ray
absorbing column densities, smaller angular diameters and greater distances
are no doubt principal culprits militating against finding PWNe in remnants
toward the inner Galaxy, as are dispersion smearing, multipath scattering, and
the high surface brightness of the radio emission from the Galactic plane
when conducting radio pulsar searches. The same factors bias the latitude 
distribution of the
detected fraction: 13\% for $|b|<0.3^{\circ}$ versus 31\% for 
$0.3^{\circ}<|b|<1.0^{\circ}$. The fraction at higher latitudes drops to
24\%, but this may in part be due to the larger fraction of SN Ia's found 
farther from the plane of the Galaxy. 

The angular size dependence shows a possible age effect: 35\% of remnants with
a diameter $D<5^{\prime}$ host a neutron star, whereas for remnants in the range
$10^{\prime}<D<20^{\prime}$ the fraction falls to 13\%. The fraction rises again
to 25--30\% for very large diameter remnants, most likely because these are
nearby objects in which very faint pulsars and/or synchrotron nebulae can be
detected.

In the most recent SNR catalog (Green 2000), 225 remnants are listed; a handful
of additions have been suggested at this meeting.
According to the best estimates of Galactic SN rates (van den Bergh \& Tammann
1991), roughly 15\% of all explosions are expected to be of Type Ia (and
thus, in most, but not all scenarios, are not expected to leave a compact
remnant). We can expect Type Ia remnants to be overrepresented in the
current sample owing to their larger scale height; it is easier to
detect remnants away from the tangle of radio-emitting H\,{\sc ii} 
regions and diffuse
non-thermal emission in the Galactic plane. Thus, we estimate that at least
40--50 remnants in the current compilation result from Type Ia
explosions. In addition, it must be the case that {\it some} fraction of
core-collapse explosions produce black holes, since we have extremely
strong evidence for the existence of stellar-mass black holes in more than
a dozen binary systems. We use ten as a conservative placeholder for the number
of currently catalogued remnants which produced black holes. This leaves
$\sim 170$ remnants in which we should expect to find a neutron star
remnant. Table 1 shows that we already have such evidence in 48 cases or nearly
one-third of the total, while Table 2 offers evidence that our current
surveys are incomplete by at least a factor of two. 

Thus, contrary to the situation just three years ago (Helfand 1998), we 
argue that there is no need to invoke either a large black hole production rate
or the creation of invisible neutron stars in order to explain the statistics of
neutron star supernova remnant associations.

\begin{table}[h]
\tiny
\caption{The current neutron star/ supernova remnant census}
\centerline{
\rotate[l]{
\begin{tabular}{|l|c|c|c|}\tableline
  & & & \\
Evidence & Plerionic Remnant & Composite Remnant & Pure Shell Remnant \\
  & & & \\\tableline
  & & & \\
Pulsar + & G106.6+2.9 [PSR J2229+6114]        & G5.4$-$1.2 (Duck) [PSR B1757$-$24] & G180.0$-$1.7 (S147) [PSR J0538+2817]\\
Supernova& G130.7+3.1 (3C~58) [PSR J0205+64]   & G11.2$-$0.3 [PSR J1811$-$1925] & G292.2$-$0.5 [PSR J1119$-$6127]    \\
Remnant  & G184.6$-$5.8 (Crab) [PSR B0531+21] & G29.7$-$0.3 (Kes 75) [PSR J1846$-$0258] & \\
(15+2)     &                                    & G34.7$-$0.4 (W44) [PSR B1853+01] &\\
         & N157B (in LMC) [PSR J0537$-$6917]  & G69.0+2.7 (CTB 80) [PSR B1951+32] &\\
         &                                    & G114.3+0.3 [PSR B2334+61] &\\
         &                                    & G263.9$-$3.3 (Vela) [PSR B0833$-$45] &\\
         &                                    & G292.0+1.8 [PSR J1124$-$5916] &\\
         &                                    & G308.8$-$0.1 [PSR J1341$-$6220] &\\
         &                                    & G320.4$-$1.2 (MSH 15$-$5{\it 2}) [PSR B1509$-$58] &\\
         &                                    &                                 & \\
         &                                    & N158A (in LMC) [PSR B0540$-$69] &\\
  & & & \\
\tableline
 & & & \\
Exotic/Possible NS   &  G54.1+0.3 [CXOU J193030.1+185214] & G0.9+0.1 [SAX J1747$-$2809]   &  G27.4+0.0 (Kes 73) [AX J1841$-$045] (AXP) \\
+ Supernova &  & G119.5+10.2 (CTA 1) [RX J000702+7302.9]  &  G29.6+0.1 [AX J1845$-$0258] (AXP?)  \\
Remnant     &  & G189.1+3.0  (IC~443) [CXOU J061705.3+222127] &  G39.7$-$2.0 [SS 433] (binary) \\
(16+1)        &  & G291.0$-$0.1 (MSH 11$-$6{\it 2}) [AX J1111$-$6040] &  G78.2+2.1 (gamma Cygni) [RX J2020.2+4026] (NS?) \\
            &  &  &  G109.1$-$1.0 (CTB 109) [1E 2259+586] (AXP) \\
            &  &  &  G111.7$-$2.1 (Cas A) [CXO J232327.9+584842] (NS?) \\
            &  &  &  G260.4$-$3.4 (Puppis A) [RX J0822$-$4300] (NS?) \\
            &  &  &  G266.2$-$1.2 (RX J0852.0$-$4622) [SAX J0852.0$-$4615] (NS?) \\
            &  &  &  G296.5+10.0 (PKS~1209--51/52) [1E 1207.4$-$5209] \\
            &  &  &  G321.9$-$0.3 [Cir X-1] (binary) \\
            &  &  &  G332.4$-$0.4 [RCW 103] (1E 161348$-$5055) (NS?) \\
            &  &  &  \\
            &  &  &  N49 (in LMC) [SGR 0526$-$66] (SGR) \\
  & & & \\
\tableline
 & & & \\
X-ray and   & G20.0$-$0.2  & G16.7+0.1 & \\
Radio nebula& G21.5$-$0.9  & G39.2$-$0.3 & \\
(9)        & G74.9+1.2    & G326.3$-$1.8 (MSH 15$-$5{\it 6}) &\\
            & G328.4+0.2   & G327.1$-$1.1 & \\
            &              & G344.7$-$0.1 & \\
  & & & \\
\tableline
 & & & \\
Radio nebula& G6.1+1.2  & G24.7+0.6   &\\
only        & G27.8+0.6 & G293.8+0.6  &\\
(8)         & G63.7+1.1 & G318.9+0.4  &\\
            &           & G322.5$-$0.1&\\
            &           & G351.2+0.1  &\\
  & & & \\
\tableline\tableline
\end{tabular}
}}
\normalsize
\end{table}

\begin{table}
\caption{The distributions of SNRs containing evidence of a young neutron star}
\begin{tabular}{ccccccc}
\multicolumn{7}{c}{\bf Galactic Longitude} \\\tableline
& $-20^{\circ} \!\!< \!\! l \!\! < \! 20^{\circ}\!$ &   $20^{\circ} \!\! < \!\! l \!\! < \! 60^{\circ}\! \!$ &   $60^{\circ} \!\! < \!\! l \!\! < \! 180^{\circ}\!$ &  $180^{\circ} \!\! < \!\! l \!\! < \! 300^{\circ}\!$ &  $300^{\circ} \!\! < \!\! l \!\! < \! 340^{\circ}\!$ & \\ 
& 7/69  &     11/43 &     11/39 &   10/30 &    9/45 \\\tableline \\
\multicolumn{7}{c}{\bf Galactic Latitude} \\\tableline
 & &      $|b|<0.3^{\circ}$ &   $0.3^{\circ}<|b|<1.0^{\circ}$ &    $|b|>1.0^{\circ}$ & & \\
 & &	 11/85  &    15/49 &       22/92 & & \\\tableline \\
\multicolumn{7}{c}{\bf Angular Size}\\\tableline
$<5'$  & $5'-10'$  & $10'-20'$  & $20'-40'$ & $40'-80'$  & $80'-160'$ & $>160'$ \\
7/20  & 12/40  &    7/53  &   10/60 &   6/33 &    4/14 &     2/6 \\\tableline\tableline
\end{tabular}
\end{table}

\section{Conclusions}

As with any maturing astronomical field (or in life, for that matter), the
simple pictures of youth grow with time more complicated and more 
frustrating --- or richer, depending on one's personal gestalt. The triumphalist picture of science as proceeding from an inspired insight (Baade
and Zwicky 1934) to a more detailed model (Pacini 1967) to an experimental
confirmation (Hewish et al.\ 1968) has always been a trifle simplistic.
Supernovae do not {\it always} ``represent the transition from an ordinary star
into a neutron star'', and the ones that do evidently provide us
with a host of possible manifestations. We have presented here a snapshot of
our current state of knowledge concerning the genesis of neutron stars.
The fact the we cite no fewer than 39 references from within the past
twenty-four months suggests the subject is a vibrant one which is benefiting
from significant improvements in observational sensitivity, primarily in the
X-ray and radio bands. It took thirty years to find the first ten secure
pulsar-remnant associations and less than three years to find the next five.
The large number of candidates discussed above assures continuing rapid 
progress. The longstanding embarrassment of all those empty supernova remnants
appears to be significantly ameliorated and is replaced by the challenge of overcoming the selection effects which hamper discoveries, and understanding
the many ways in which young neutron stars manifest their presence.  Such
understanding might even lead to quantitative constraints on the conditions
present in core-collapse explosions and the resulting nucleosynthetic yields
which determine the chemical composition of the Galaxy.

\acknowledgments

V.M.K. is a Sloan Fellow
and CRC Chair.  This work is supported by SAO grant GO0-1133A, and by
NASA LTSA grant NAG5-8063 and NSERC grant Rgpin 228738-00
to V.M.K. and by SAO G00-1130X to D.J.H.

\end{document}